\documentclass{ifacconf}

\usepackage{graphicx}      
\usepackage{natbib}        
\usepackage{amsmath}
\usepackage{amssymb}
\usepackage{graphicx}
\usepackage{bm}
\usepackage{xcolor}

\newcommand{\Transp}{\intercal}

\newcommand{\jointAxisINF}{\bm{l}_\mathrm{i}}
\newcommand{\jointAxisKNF}{\bm{l}_\mathrm{k}}
\newcommand{\jointAxisI}[1]{\bm{l}_\mathrm{i}^\mathrm{b_#1}}
\newcommand{\jointAxisK}[1]{\bm{l}_\mathrm{k}^\mathrm{b_#1}}
\newcommand{\jointAxisPer}{\bm{l}_{\perp}}
\newcommand{\jointAxisPerR}{\bm{l}_{\perp}^\mathrm{r}}
\newcommand{\jointAxisPerJ}{\bm{l}_{\perp}^\mathrm{b_j}}

\newcommand{\cRIn}{\bm{R}_\mathrm{b_i}^\mathrm{r}}
\newcommand{\cRJn}{\bm{R}_\mathrm{b_j}^\mathrm{r}}
\newcommand{\cRKn}{\bm{R}_\mathrm{b_k}^\mathrm{r}}

\newcommand{\cRIndot}{\dot{\bm{R}}_\mathrm{b_i}^\mathrm{r}}
\newcommand{\cRJndot}{\dot{\bm{R}}_\mathrm{b_j}^\mathrm{r}}
\newcommand{\cRKndot}{\dot{\bm{R}}_\mathrm{b_k}^\mathrm{r}}

\newcommand{\cYgyrI}{\bm{y}_\mathrm{i,\omega}^\mathrm{b_i}}
\newcommand{\cYgyrJ}{\bm{y}_\mathrm{j,\omega}^\mathrm{b_j}}
\newcommand{\cYgyrK}{\bm{y}_\mathrm{k,\omega}^\mathrm{b_k}}
\newcommand{\cYgyrIdot}{\dot{\bm{y}}_\mathrm{i,\omega}^\mathrm{b_i}}

\newcommand{\cYgyrKdot}{\dot{\bm{y}}_\mathrm{k,\omega}^\mathrm{b_k}}
\newcommand{\cYgyrImeas}{\tilde{\bm{y}}_\mathrm{i,\omega}^\mathrm{b_i}}

\newcommand{\cYgyrKmeas}{\tilde{\bm{y}}_\mathrm{k,\omega}^\mathrm{b_k}}
\newcommand{\cYgyr}{\bm{y}_\mathrm{\omega}^\mathrm{b}}

\newcommand{\crossP}[1]{\left[ {#1} \right]_\times}

\newcommand{\angveli}{\bm{\omega}_\mathrm{i}^\mathrm{b_i}}

\newcommand{\angvelk}{\bm{\omega}_\mathrm{k}^\mathrm{b_k}}
\newcommand{\angvelkN}{\bm{\omega}_\mathrm{k}^\mathrm{r}}
\newcommand{\angveliN}{\bm{\omega}_\mathrm{i}^\mathrm{r}}

\newcommand{\angvelJI}{\bm{\omega}_{\mathrm{\bm{l}_i}}^\mathrm{b_i}}
\newcommand{\angvelJK}{\bm{\omega}_{\mathrm{\bm{l}_k}}^\mathrm{b_k}}
\newcommand{\angvelParJ}{\bm{\omega}_{\bm{l}_{\perp}}^\mathrm{b_j}}
\newcommand{\angvelParJdot}{\dot{\bm{\omega}}_{\bm{l}_{\perp}}^\mathrm{b_j}}
\newcommand{\angvelResJ}{\bm{\omega}_{\bm{l}_{\not\perp}}^\mathrm{b_j}}

\newcommand{\cRIK}{\bm{R}^\mathrm{b_i}_\mathrm{b_k}}
\newcommand{\cRIJ}{\bm{R}^\mathrm{b_i}_\mathrm{b_j}}
\newcommand{\cRKJ}{\bm{R}^\mathrm{b_k}_\mathrm{b_j}}
\newcommand{\cRIJdot}{\dot{\bm{R}}^\mathrm{b_i}_\mathrm{b_j}}
\newcommand{\cRKJdot}{\dot{\bm{R}}^\mathrm{b_k}_\mathrm{b_j}}

\newcommand{\biasGyr}{\bm{b}^\mathrm{b}_\mathrm{\omega}}
\newcommand{\noiseGyr}{\bm{e}^\mathrm{b}_\mathrm{\omega}}

\newcommand{\biasGyrI}{\bm{b}_\mathrm{i,\omega}^\mathrm{b_i}}

\newcommand{\biasGyrK}{\bm{b}_\mathrm{k,\omega}^\mathrm{b_k}}
\newcommand{\noiseGyrI}{\bm{e}_\mathrm{i,\omega}^\mathrm{b_i}}

\newcommand{\noiseGyrK}{\bm{e}_\mathrm{k,\omega}^\mathrm{b_k}}

\newcommand{\angvelParJP}{\breve{\omega}_{l_{\perp}}}
\newcommand{\angvelResJP}{\breve{\omega}_{l_{\not\perp}}}
\newcommand{\angvelJIP}{\breve{\omega}_{\mathrm{l_i}}}
\newcommand{\angvelJKP}{\breve{\omega}_{\mathrm{l_k}}}

\newcommand{\quatI}{\bm{q}_\mathrm{b_i}^\mathrm{r}}
\newcommand{\quatJ}{\bm{q}_\mathrm{b_j}^\mathrm{r}}
\newcommand{\quatK}{\bm{q}_\mathrm{b_k}^\mathrm{r}}
\newcommand{\quatRel}{\bm{q}_\mathrm{b_j}^\mathrm{b_i}}
\newcommand{\quatRelK}{\bm{q}_\mathrm{b_k}^\mathrm{b_i}}

\newcommand{\quatRelJErrAng}{\varphi_\mathrm{err}^\mathrm{ji}}
\newcommand{\quatRelKErrAng}{\varphi_\mathrm{err}^\mathrm{ki}}
\newcommand{\quatFromAngvel}[1]{\bm{q}\left({#1}\right)}
\newcommand{\quatRot}[2]{\left[{#1}\right]_{{#2}(t)}}

\newcommand{\Ts}{T_\mathrm{s}}

\newcommand{\norm}[1]{\left\| {#1} \right\|_2}

\newcommand{\Unit}[1]{\left[\mathrm{#1}\right]}

\DeclareMathOperator*\argmin{arg \, min \,}

\begin{document}
\begin{frontmatter}

\title{Sparse Magnetometer-free Inertial Motion Tracking -- A Condition for Observability in Double Hinge Joint Systems} 


\author[First]{Karsten Eckhoff} 
\author[Second]{Manon Kok} 
\author[Third]{Sergio Lucia}
\author[First]{Thomas Seel}

\address[First]{Control Systems Group, Technische Universit\"at Berlin, 10587 Berlin, Germany (e-mail: seel@control.tu-berlin.de)}
\address[Second]{Delft Center for Systems \& Control, Delft University of Technology, 2628 CD Delft, The Netherlands (e-mail: m.kok-1@tudelft.nl)}
\address[Third]{Internet of Things for Smart Buildings, Technische Universit\"at Berlin, 10587 Berlin, Germany (e-mail: sergio.lucia@tu-berlin.de)}

\begin{abstract}
Inertial measurement units are commonly used in a growing number of application fields to track or capture motions of kinematic chains, such as human limbs, exoskeletons or robotic actuators. A major challenge is the presence of magnetic disturbances that result in unreliable magnetometer readings. Recent research revealed that this problem can be overcome by exploitation of kinematic constraints. While typically each segment of the kinematic chain is equipped with an IMU, a novel approach called sparse inertial motion tracking aims at infering the complete motion states from measurements of a reduced set of sensors. 
In the present contribution, we combine the magnetometer-free and the sparse approach for real-time motion tracking of double-hinge joint systems with non-parallel joint axes. Analyzing the observability of the system, we find a condition which assures that the relative orientations between all segments are uniquely determined by a kinematic constraint, which contains only the gyroscope readings. Furthermore, we propose a moving-horizon estimator and validate it in a simulation study of three movements with different degrees of excitation. The results of this study confirm all theoretical conjectures and demonstrate that magnetometer-free sparse inertial real-time motion tracking is feasible under precise and simple excitation conditions.
\end{abstract}


\end{frontmatter}

\section{Introduction}
Inertial measurement units (IMUs) are used in several application domains mainly for the purpose of localization, motion capture and real-time motion tracking. The object of interest often consists of multiple segments that are connected by joints and form a kinematic chain, for example human limbs or robot manipulators. 
Especially in health applications, wearable IMUs are commonly used nowadays~\citep{Wong:2015, Buke:2015}. \par
Typically, one IMU is attached to each segment of the kinematic chain \citep{kokHS:2014, Miezal:2016} to be analyzed. However, hardware cost, unobtrusiveness and donning time could be improved if the motion of all segments could be captured or tracked without equipping all segments with an IMU. This approach to use a reduced number of sensors is known as sparse inertial motion tracking. In~\cite{Marcard:2017} only 6 IMUs are used to capture arm, leg, trunk and head movements. They use an offline optimization framework to fit sequences of orientation and acceleration data to the pose of a statistical body model. Another example is~\cite{Huang:2018} where a deep neural network is used for real-time estimation of the body pose from the measurements of, again, only 6 IMUs. Both works use an extensive model to overcome the ambiguity that multiple poses generate the same sensor readings, but they do not provide an analysis of the question under which circumstances or conditions it is possible to uniquely determine the motion states. Finally, both methods rely on magnetometer readings. \par
A major problem in inertial motion tracking is the presence of distortions and disturbances of the local magnetic field, which occur near ferro-magnetic material or electronic devices and especially in indoor environments. This means that, in any of the mentioned circumstances, the heading information, i.e. the orientation around the vertical axis, cannot be reliably inferred from the magnetometer readings. While the absolute heading of the entire kinematic chain is often less relevant or simply known by construction, the relative heading between the segments is crucial for determining the pose of the kinematic chain as well as relative motion parameters such as joint angles. \par Researchers have proposed several approaches that infer the missing relative heading information by exploiting kinematic constraints in different types of joints and kinematic chains. In~\cite{laidigSS:2017} a quaternion-based method is proposed to determine the joint angle of a hinge joint in real time. The approach does not rely on magnetometer readings and is evaluated by simulation. A similar method for the case of 2D-joints is proposed and evaluated experimentally in~\cite{laidigLS:2019}. An offline optimization-based approach to magnetometer-free inertial motion capture is presented and used to estimate the pose of the lower body in~\cite{kokHS:2014}. While all authors agree that some minimum level of excitation must be present or remaining near singular poses should be avoided, none of these works provides precise conditions that the movement must fulfill to assure that the motion states can be determined uniquely from the measurements and constraints. \par
Exploiting kinematic constraints to track the motion states in real time typically requires a method for solving a constrained optimization problem at runtime. A rather powerful tool, which has not been used very frequently for inertial motion tracking, is Moving Horizon Estimation (MHE). MHE can be applied to non-linear systems, and it can handle different kinds of constraints including state constraints. In the past, MHE has been used successfully to estimate the position, velocity and orientation of an airplane from a global navigation satellite system receiver and an IMU~\citep{Girrbach:2017}.\par
To the best of our knowledge there is no previous work that achieves sparse \emph{and} magnetometer-free inertial motion tracking and no previous work that investigates conditions on the motion that assure observability in magnetometer-free or sparse inertial motion tracking. We consider double-hinge joint systems with non-parallel axes, propose a state space model and investigate conditions for observability of the motion states given sparse gyroscope readings. Furthermore, we propose a MHE method that solves the online estimation problem. It is tested in a simulation study for three motion scenarios with different levels of excitation.

\section{System Model}

\subsection{Kinematic Model}
\begin{figure}[!ht] 
\centering 
\includegraphics[width=0.48\textwidth]{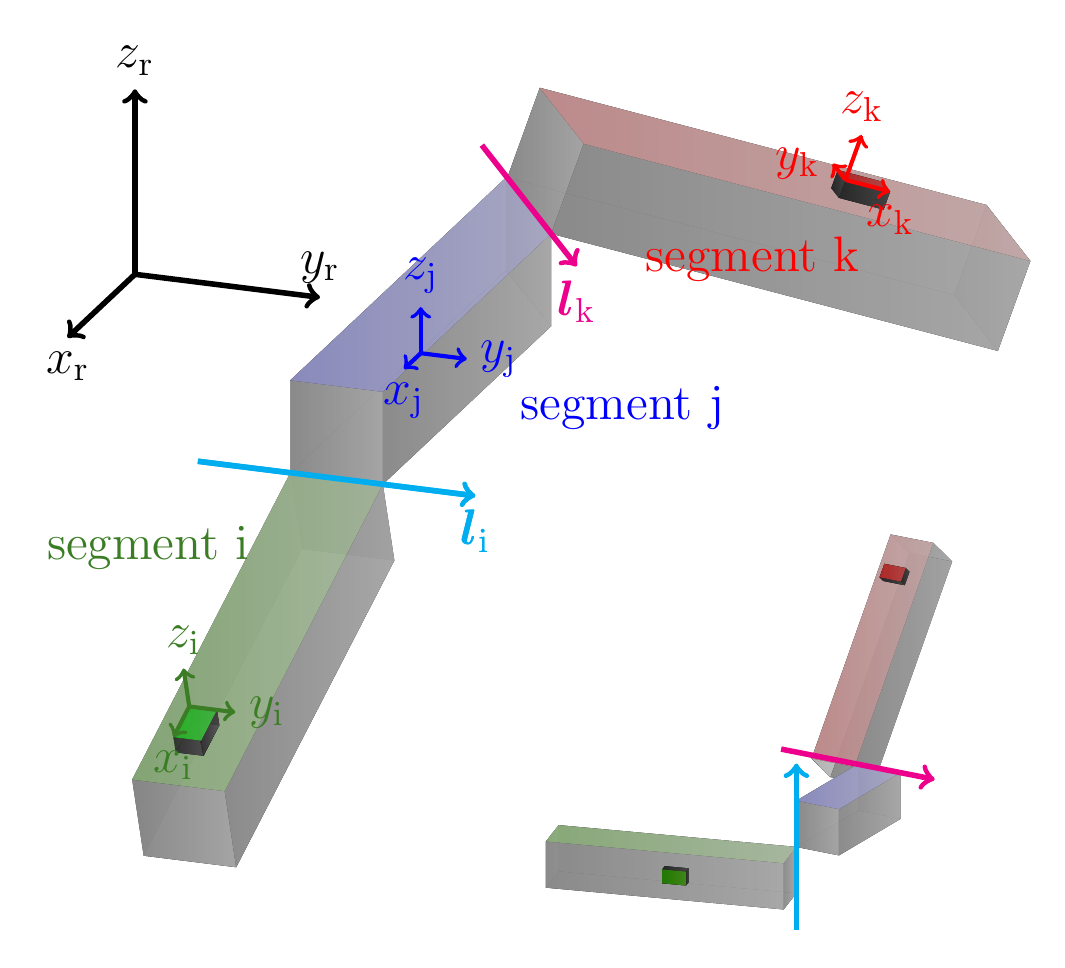}
\caption{Two examples of a kinematic chain consisting of three segments (i, j and k) connected in series by two hinge joints with non-parallel joint axes $\jointAxisINF$ and $\jointAxisKNF$.}
\label{fig:coordSysHJ}
\end{figure}

Consider a kinematic chain consisting of three segments (segment i, segment j and segment k) connected in series by hinge joints.
Let the middle segment of the chain be segment j, which is connected to segment i via the joint axis $\jointAxisINF$ and to segment k via the joint axis $\jointAxisKNF$.
Consider the case in which the two joint axes $\jointAxisINF$ and $\jointAxisKNF$ are non-parallel.
Fig.~\ref{fig:coordSysHJ} shows two examples of such a kinematic chain.
Note that the described structure could also be a part of a longer kinematic chain. \par
An IMU is attached to each of the outer segments (segment i and segment k). Without loss of generality, the coordinate systems of the sensor and the corresponding segment are assumed to be identical. There exist a number of methods for aligning the coordinate system of a segment with the coordinate system of the attached IMU, see e.g.~\cite{laidigMS:2017,taetzBM:2016,kokHS:2014,Olsson2019_FUSION,Graurock2016_IFESS}. 
Henceforth the coordinate system of the segment is referred to as the b-frame (body-frame), e.g. the b-frame of segment i is denoted by $\mathrm{b}_i$.
The coordinates of the joint axes expressed in the b-frames of the adjacent segments are denoted by $\jointAxisI{i}$, $\jointAxisI{j}$, $\jointAxisK{j}$ and $\jointAxisK{k}$. These parameters are considered to be known. 
The motion of the kinematic chain is fully determined by the translational motion of segment j in some inertial reference frame, henceforth referred to as r-frame, and by the rotations of the three segments with respect to that same reference frame. In the following, we let the translation be arbitrary and only consider the rotational motion. 
The orientations of all three segments are parametrized by rotation matrices $\cRIn$, $\cRJn$ and $\cRKn$, but all arguments and results hold likewise for quaternion representations.  \par
Since the segments are connected by hinge joints, the movement of the three segments is constrained. This can be described by the following two constraints:
\begin{subequations}
\begin{align}
\cRIn \jointAxisI{i} &= \cRJn \jointAxisI{j},
\\
\cRJn \jointAxisK{j} &= \cRKn \jointAxisK{k}.
\end{align}
\label{eq:HJconst}
\end{subequations}
The constraints \eqref{eq:HJconst} can be interpreted in the following way: The coordinates of the joint axis (expressed in their corresponding b-frame) must be identical when they are transformed into the same frame. Here they are both transformed into the r-frame.
Furthermore, it has been shown that the following constraint can be exploited for the case of 2D-joints~\citep{laidigMS:2017}:
\begin{align}
& {\angveliN}^\Transp \left( \cRIn \jointAxisI{i} \times \cRKn \jointAxisK{k} \right) = {\angvelkN}^\Transp \underbrace{\left( \cRIn \jointAxisI{i} \times \cRKn \jointAxisK{k} \right)}_{\jointAxisPerR},
\label{eq:HJ2Dconst}
\end{align}
where $\angveliN$ and $\angvelkN$ denote the angular velocities of segments i and k, respectively, both expressed in the r-frame.
The constraint~\eqref{eq:HJ2Dconst} describes that the projection of the angular velocity onto the axis $\jointAxisPer$, which is perpendicular to both joint axes, has to be equal for segment i and segment k.
Note that there are no limitations regarding the range of motion of the joints which would lead to additional constraints but would also be more restrictive. 

\subsection{Measurement Model}
Consider only the gyroscope readings of the IMUs, i.e. the measurements $\cYgyrI$ and $\cYgyrK$ of the angular velocities $\angveli$ and $\angvelk$ of the segments i and k: 
\begin{subequations}
\begin{align}
\cYgyrI &= \angveli + \biasGyrI + \noiseGyrI,
\\
\cYgyrK &= \angvelk + \biasGyrK + \noiseGyrK.
\end{align}
\label{eq:measmodgyr}
\end{subequations}
The measurements are corrupted by a bias error $\biasGyr$ and measurement noise $\noiseGyr$.
For the theoretical analysis, the sensors are considered to be ideal in the sense that $\biasGyr$ and $\noiseGyr$ are zero. In the simulation study, non-zero biases and measurement noise will be considered.

\section{Observability Analysis}
In this section, we investigate under which conditions on a movement of the kinematic chain it is possible to uniquely determine the orientations of the three segments given the inertial measurements of the outer segments.
To this end, we describe the rotational motion of the system by a state space model, introduce the concept of instantaneous observability under partially unknown input and study conditions that assure this property.

\subsection{System Dynamics}
We model the dynamics of the orientation of the three segments as~\citep{kokHS:2017}
\begin{subequations}
\begin{align}
\cRIndot &= \cRIn \crossP{\cYgyrI},
\\
\cRJndot &= \cRJn \crossP{\cYgyrJ},
\\
\cRKndot &= \cRKn \crossP{\cYgyrK}.
\end{align}
\label{eq:HJsys_dyn}
\end{subequations}
Here, $\crossP{\cYgyr}$ is a skew-symmetric matrix which represents the cross product matrix of the angular velocity $\cYgyr$.
While $\cYgyrI$ and $\cYgyrK$ are known measurements, segment k is not equipped with an IMU and $\cYgyrJ$ is thus unknown for all time instances.
The constraints \eqref{eq:HJconst} and \eqref{eq:HJ2Dconst} are formulated as virtual system outputs
\begin{subequations}
\begin{align}
\bm{y}_1 &= \cRIn \jointAxisI{i} - \cRJn \jointAxisI{j} ,\label{eq:HJconstIJ}
\\
\bm{y}_2 &= \cRJn \jointAxisK{j} - \cRKn \jointAxisK{k} ,\label{eq:HJconstIK}
\\
y_3 &= \left( \cRIn \cYgyrI - \cRKn \cYgyrK \right)^\Transp \left( \cRIn \jointAxisI{i} \times \cRKn \jointAxisK{k} \right) ,\label{eq:HJconst2d}
\end{align}
\label{eq:HJOutputs}
\end{subequations}
all of which are known to be constantly zero.

\subsection{Conditions on the movement of the kinematic chain for observability of the orientations}
In the literature, there exist different notions of observability for nonlinear systems. One that seems particularly useful for our purpose is called \textit{local observability} by G. Besan\c{c}on~\citep{besancon:2007}. Here, the notion of \emph{local} means that it is possible to determine the states from the measurements and inputs of an arbitrary small time frame. On that basis we define an observability property called \textit{instantaneous observability under partially unknown input}.

Consider a general nonlinear system of the form
\begin{align}
\begin{split}
\dot{\bm{x}} &= \bm{f}(\bm{x}(t),\,\bm{u}(t)),\\
 \begin{bmatrix}
y_1, y_2, \ldots, y_p
\end{bmatrix}^\Transp = \bm{y} &= \bm{h}(\bm{x}(t),\,\bar{\bm{u}}(t)),
\end{split}
\label{eq:obssys}
\end{align}
where $\bm{x}\in\mathbb{R}^n$ denotes the state vector and $\bm{y} \in \mathbb{R}^p$ the vector of the measurements. 
Futhermore, $\bm{u} \in \mathbb{R}^m$ denotes the vector of the inputs and $\bar{\bm{u}}$ the subset of $\bm{u}$ that contains the known inputs. Given $p$ orders $N_1, N_2,\ldots, N_p \geq 0$ with $\hat{N}:=\sum^p_{i=1}\left(N_i+1\right)$, a time instant $t$ and some known input $\bar{\bm{u}}(t)$, define the mapping $\Phi_{\bar{\bm{u}},N_1,\ldots,N_p}:\,\mathbb{R}^n\to\mathbb{R}^{\hat{N}}$ by
\begin{align}
\Phi_{\bar{\bm{u}},N_1,\ldots,N_p}\left(\bm{x}(t)\right):=\begin{bmatrix}
[y_1(t),
\dot{y}_1(t),
\ldots,
y_1^{(N_1)}(t)]^\Transp \\
[y_2(t),
\dot{y}_2(t),
\ldots,
y_2^{(N_2)}(t)]^\Transp \\
\vdots\\
[y_p(t),
\dot{y}_p(t),
\ldots,
y_p^{(N_p)}(t)]^\Transp
\end{bmatrix},\label{eq:outmapp}
\end{align}%
where $y^{(N)}$ denotes the $N$-th time-derivative of $y$ and all orders $N_1, N_2,\ldots, N_p$ must be chosen such that the mapping does not depend on any of the unknown inputs.
\begin{defn}
For a given known input $\bar{\bm{u}}$ at time $t$, the system \eqref{eq:obssys} is called \emph{instantaneously observable under partially unknown input} if there exist $N_1,N_2,\ldots,N_p \geq 0$ such that the mapping $\Phi_{\bar{\bm{u}},N_1,\ldots,N_p}\left(\bm{x}(t)\right)$ is injective, i.e. there are no two points in state space that yield the same vector of output derivatives.
\end{defn}
Note that this definition describes the ideal case in which all states can be inferred. For magnetometer-free inertial motion tracking, it is well known that the overall heading of the entire kinematic chain cannot be determined, since the horizontal axes of the inertial frame of reference point into arbitrary directions. However, the magnetometer-free motion tracking problem becomes well posed if the orientation of one of the outer segments is assumed to be known, which is equivalent to choosing a specific reference frame. In that case, the task is to determine the orientations of all remaining segments. 
\begin{thm} The system \eqref{eq:HJsys_dyn}, \eqref{eq:HJOutputs} is instantaneously observable under partially unknown input for any time instant $t$ for which the following two conditions hold:
\begin{enumerate}
\item The orientation of segment i or segment k is known.
\item The angular velocity of segment j is neither parallel to the axis $\jointAxisPer$ nor perpendicular to it. 
\end{enumerate}
\label{th:obsth}
\end{thm}
The proof is given in Appendix \ref{app:proof}. \par
Note that the constraints \eqref{eq:HJOutputs}, which serve as virtual outputs of the system \eqref{eq:HJsys_dyn}, do not contain the unknown gyroscope measurements $\cYgyrJ$ of segment j.

\begin{cor}
If we are only interested in the \emph{relative} orientations between all segments, we can choose any arbitrary orientation for one of the outer segments and then apply Theorem~\ref{th:obsth}. This implies that, if none of the orientation are known, then it is still possible to determine the relative orientation between the three segments.
\label{cor:obs}
\end{cor}

\begin{rem}
It follows from the proof that if the angular velocity of segment j is constantly perpendicular to $\jointAxisPer$, then all higher derivatives of the constraint \eqref{eq:HJconst2d} are also $\bm{0}$. This represents a degenerate case that corresponds to the degenerate case of Lemma~\ref{lem:wahba}, and the (relative) orientations are not unique in the defined sense.
\label{rm:notobs}
\end{rem}

\begin{rem}
 Note that all results hold independent of the translational motion of the middle segment. Only the rotational motion of the segments influences the considered observability properties.
\end{rem}

\section{MHE Formulation}
In the previous chapter we showed under which conditions on the movement of the kinematic chain there exists only one unique motion state that fulfills the constraints. We now propose a MHE approach that can be used to determine the motion states from the gyroscope readings and the kinematic constraints.\par
The MHE implementation is based on the python-based \mbox{do-mpc} framework~\citep{LuciaTSE:2017}, which allows modularized implementation and testing support for optimal control schemes based on Model Predictive Control approaches. Furthermore, the framework CasADi~\citep{Anderson2018} in combination with IPOPT~\citep{WaechterB:2006} is used to solve the non-linear optimization problem and as linear solver ma27 from HSL~\citep{HSL} is used. \par
To obtain an efficient implementation of the MHE, unit quaternions (denoted by $\bm{q}$) are used to parametrize orientations. The symbol $\otimes$ is used to denote the quaternion multiplication. The transformation of a measurement $\bm{o}$ from the body frame into the navigation frame is defined in the following way:
\begin{align*}
\bm{o}^\mathrm{r} &= \quatRot{\bm{o}^\mathrm{b}}{q_\mathrm{b}^\mathrm{r}} = q^\mathrm{r}_\mathrm{b} \otimes \begin{bmatrix}0\\ \bm{o}^\mathrm{b}\end{bmatrix} \otimes \left( q^\mathrm{r}_\mathrm{b} \right) ^\star,
\end{align*}
where $\left(\bm{q}\right)^\star$ denotes the conjugate of a quaternion. Furthermore, the mapping from an angular velocity $\bm{\omega}$ to a unit quaternion $\bm{q}$ is defined by
\begin{align*}
\quatFromAngvel{\bm{\omega}} &= \begin{bmatrix}
\cos \left( \frac{\alpha}{2} \right)\\
\hat{\bm{\omega}}\sin \left( \frac{\alpha}{2} \right)
\end{bmatrix},
\quad \alpha=\frac{\norm{\bm{\omega}}}{\Ts}, \quad \hat{\bm{\omega}}=\frac{\bm{\omega}}{\norm{\bm{\omega}}},
\end{align*} where $T_s$ denotes the sample time.

\subsubsection{Optimization Problem}
The optimization problem that the MHE solves at every sampling instant is given by  
\begin{small}
\begin{align}
\begin{array}{cll}
\underset{\bm{x}(t_s:t_e),\,\bm{u}(t_s:t_e)}{\argmin} & a(\bm{x}(t_s),\,\bm{u}(t_s)) + \sum^{t_e}_{t=t_s}s(\bm{x}(t),\,\bm{u}(t))\\
\mathrm{subject\ to} &\bm{x}\left(t+1\right) = \bm{f}\left(\bm{x}\left(t\right),\,\bm{u}\left(t\right)\right),\\
&\norm{\quatI(t)} = 1,\\
&\norm{\quatJ(t)} = 1,\\
&\norm{\quatK(t)} = 1,
\end{array}
\label{eq:mheoptiprob}
\end{align}
\end{small}%
where $t$ is an integer-valued time variable, $t_s$ denotes the first time instance of the current time interval and $t_e$ the last one.
We obtain the discrete-time dynamics described by equation \eqref{eq:dynamicsDiscrete} if we apply Euler discretization to the continuous-time dynamics \eqref{eq:HJsys_dyn} with unit quaternions as orientation parameters~\citep{kokHS:2017}.
\begin{align}
\bm{f}\left(\bm{x}\left(t\right),\,\bm{u}\left(t\right)\right)=\begin{bmatrix} \quatI(t)\otimes\quatFromAngvel{\cYgyrI(t)}
\\
\quatJ(t)\otimes\quatFromAngvel{\cYgyrJ(t)}
\\
\quatK(t)\otimes\quatFromAngvel{\cYgyrK(t)}
\end{bmatrix}
\label{eq:dynamicsDiscrete}
\end{align}
The state vector $\bm{x}$ and the input vector $\bm{u}$ of the system are 
\begin{subequations}
\begin{align}
\bm{x} &= \begin{bmatrix}
{\quatI}^\Transp & {\quatJ}^\Transp & {\quatK}^\Transp
\end{bmatrix}^\Transp,\\
\bm{u} &= \begin{bmatrix}
{\cYgyrI}^\Transp & {\cYgyrJ}^\Transp & {\cYgyrK}^\Transp
\end{bmatrix}^\Transp,
\end{align}
\end{subequations}
and the stage cost $s$ and the arrival cost $a$ are
\begin{small}
\begin{subequations}
\begin{align}
s &= \bm{c}_1(t)^\Transp \bm{W}_{\bm{c}_1} \bm{c}_1(t) + \bm{c}_2(t)^\Transp \bm{W}_{\bm{c}_2} \bm{c}_2(t) + w_{c_3} c_3^2
\nonumber\\
& \quad + \left(\cYgyrI(t) - \cYgyrImeas(t)\right)^\Transp \bm{W}_{\cYgyrI} \left(\cYgyrI(t) - \cYgyrImeas(t)\right)
\nonumber\\
& \quad + \left(\cYgyrK(t) - \cYgyrKmeas(t)\right)^\Transp \bm{W}_{\cYgyrK} \left(\cYgyrK(t) - \cYgyrKmeas(t)\right),
\\
a &= \left(\bm{x}(t_s) - \bm{x}_\mathrm{pre}(t_s)\right)^\Transp \bm{W}_a \left(\bm{x}(t_s) - \bm{x}_\mathrm{pre}(t_s)\right),
\end{align}
\end{subequations}
\end{small}
where $\bm{c}_1$, $\bm{c}_2$ and $c_3$ are the output constraints
\begin{subequations}
\begin{align}
\bm{c}_1(t) =& \quatRot{\jointAxisI{i}}{\quatI} - \quatRot{\jointAxisI{j}}{\quatJ},
\\
\bm{c}_2(t) =& \quatRot{\jointAxisK{j}}{\quatJ} - \quatRot{\jointAxisK{k}}{\quatK},
\\
c_3(t) =& \left( \quatRot{\cYgyrI(t)}{\quatI} - \quatRot{\cYgyrK(t)}{\quatK} \right)^\Transp
\nonumber\\
& \left( \quatRot{\jointAxisI{i}}{\quatI} \times \quatRot{\jointAxisK{k}}{\quatK} \right).
\end{align}
\end{subequations}
Furthermore, $\tilde{y}_\mathrm{\omega}$ denotes the sensor readings and $x_\mathrm{pre}(t_s)$ the state estimate obtained by the previous MHE step at time $t_s$.

\subsubsection{Parameters}
A sample time of $T_s=0.01\,\mathrm{s}$ is used, the estimation horizon $H = t_e - t_s$ is $75$ samples, and the following weights are used:
\begin{equation}
\begin{gathered}
\bm{W}_{\bm{c}_1} = 2.5\cdot 10^3 \bm{I}_{3},\; \bm{W}_{\bm{c}_2} = 2.5\cdot 10^3 \bm{I}_{3},\\
w_{c_3} = 1.25\cdot 10^4,\; \bm{W}_a = 2\cdot 10^3 \bm{I}_{12},\\
\bm{W}_{\cYgyrI} = \frac{360}{2\pi}\bm{I}_{3}, \bm{W}_{\cYgyrK} = \frac{360}{2\pi}\bm{I}_{3},
\end{gathered}
\end{equation}
where $\bm{I}_m$ denotes the identity matrix of size $m$. The weights and the estimation horizon were tuned via simulation analysis.

\section{Simulation}
We now use the proposed MHE to estimate the motion states for different types of movements and investigate whether the simulation results agree with the theoretical results. For each movement we estimate the states in two different modes: first with the knowledge of the true orientation of segment i and then again without any a-priori knowledge of any of the orientations. The first mode (m1) aims at confirming Theorem~\ref{th:obsth}, while the second mode (m2) one aims at confirming Corollary~\ref{cor:obs}. 

\subsection{Kinematic Chain and Measurements}
We simulate three segments with a length of $4\,\mathrm{cm}$ where the joint(s) are located at the end(s) of the segment. The IMUs at segment i and segment k have a distance of $2\,\mathrm{cm}$ to the corresponding joint axis and the coordinates of the joint axes are:
\begin{equation}
\begin{gathered}
\jointAxisI{i} = \begin{bmatrix}
1 & 0 & 0
\end{bmatrix}^\Transp,\; \jointAxisI{j} = \begin{bmatrix}
1 & 0 & 0
\end{bmatrix}^\Transp
\\
\jointAxisK{k} = \begin{bmatrix}
1 & 0 & 0
\end{bmatrix}^\Transp,\; \jointAxisK{j} = \begin{bmatrix}
\frac{1}{\sqrt{2}} & \frac{1}{\sqrt{2}} & 0
\end{bmatrix}^\Transp
\end{gathered}
\end{equation}
The measurement noise of the sensors is set to%
\begin{small}
\begin{equation}
\begin{gathered}
\noiseGyrI \sim \mathcal{N}\left(0,\,1\Unit{\frac{{}^\circ}{s}}\right),\;
\noiseGyrK \sim \mathcal{N}\left(0,\,1\Unit{\frac{{}^\circ}{s}}\right),
\end{gathered}
\end{equation}
\end{small}%
and we simulate a bias error of
\begin{subequations}
\begin{align}
\biasGyrI &= \begin{bmatrix}
0.2 & -0.2 & 0.2
\end{bmatrix}^\Transp\Unit{\frac{{}^\circ}{s}},\\
\biasGyrK &= \begin{bmatrix}
0.2 & 0.2 & -0.2
\end{bmatrix}^\Transp\Unit{\frac{{}^\circ}{s}}.
\end{align}
\end{subequations}

\subsection{Definition of the Movements of the Kinematic Chain}

\subsubsection{Non-observable Movement (no-M):}
The segment j only rotates around the joint axis $\jointAxisI{j}$ and the position of segment j changes arbitrarily. Segment i and segment k rotate arbitrarily around their corresponding joint axis. \par
\begin{figure}[!ht]
\vspace{-2 mm}
\includegraphics[width=1\linewidth]{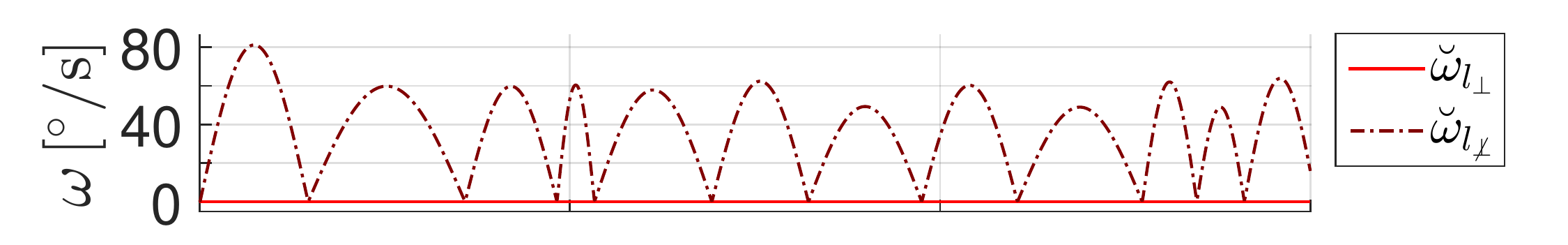} 
\includegraphics[width=1\linewidth]{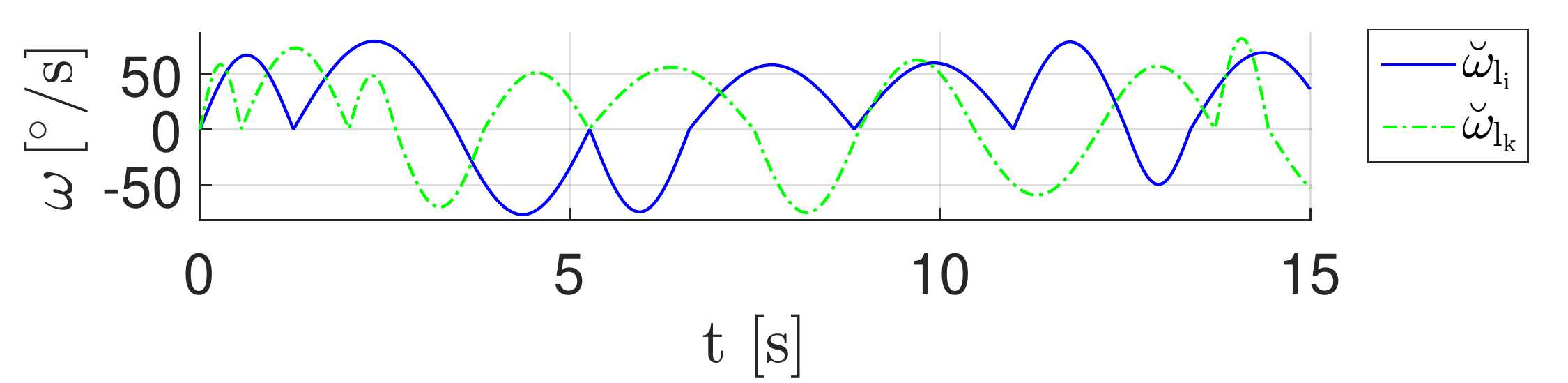}  
\caption{Angular velocities $\angvelParJ$, $\angvelResJ$, $\angvelJI$ and $\angvelJK$ projected onto their corresponding axis for the non-observable movement.}
\label{fig:noM}
\end{figure}
Fig.~\ref{fig:noM} shows the projections $\angvelParJP$, $\angvelResJP$, $\angvelJIP$ and $\angvelJKP$ of the angular velocities $\angvelParJ$, $\angvelResJ$, $\angvelJI$ and $\angvelJK$ onto their corresponding axis. Notice that $\angvelParJP$ is constantly zero.

\subsubsection{Minimal-observable Movement (mo-M):}
Segment j rotates with constant angular velocity around the axis $\begin{bmatrix}
0 & \frac{1}{2} & \frac{\sqrt{3}}{2}
\end{bmatrix}^\Transp$ which is non-parallel to all three axis $\jointAxisINF$, $\jointAxisKNF$ and $\jointAxisPer$.
The position of segment j is constant just like the joint angles between segment i and segment k. \par
\begin{figure}[!ht]
\vspace{-2 mm}
\includegraphics[width=1\linewidth]{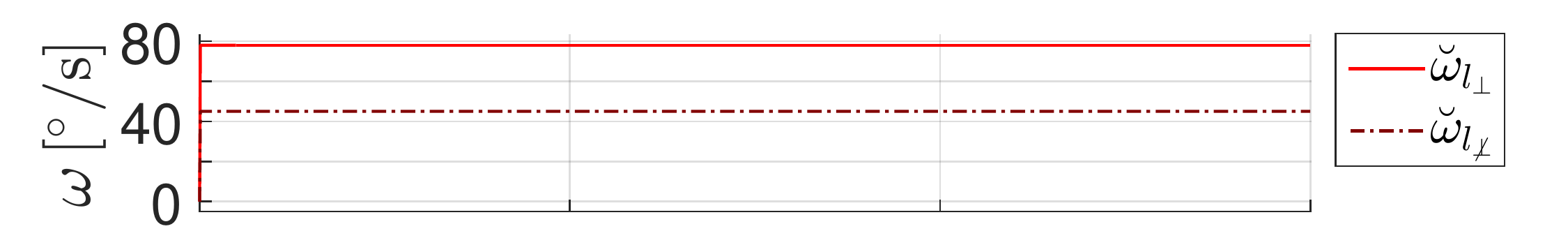} 
\includegraphics[width=1\linewidth]{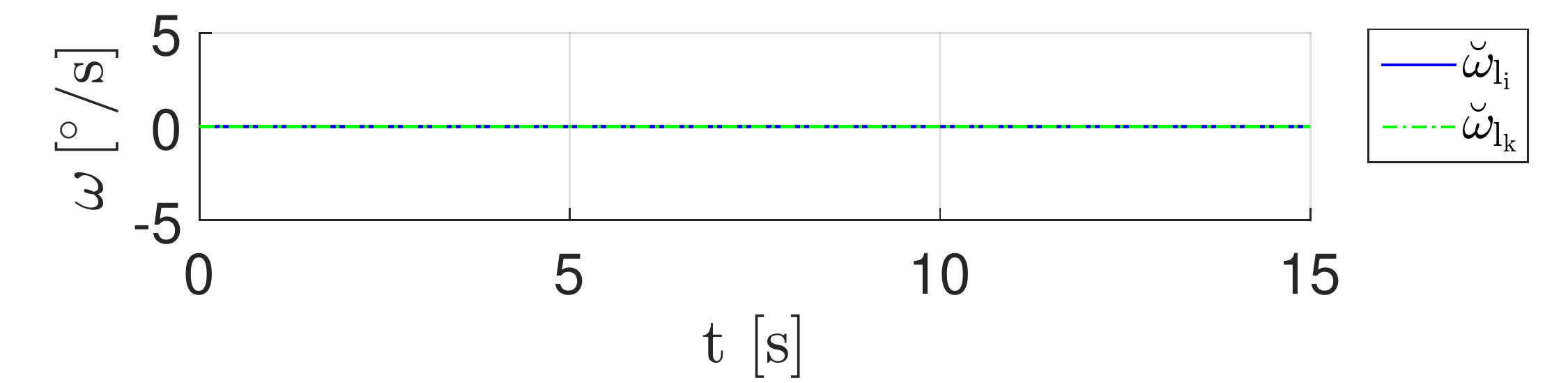}  
\caption{Angular velocities $\angvelParJ$, $\angvelResJ$, $\angvelJI$ and $\angvelJK$ projected onto their corresponding axis for the minimal-observable movement.}
\label{fig:moM}
\end{figure}
Notice that $\angvelParJP$ and $\angvelResJP$ are non-zero for all time instances.

\subsubsection{Random Movement (rd-M):}
The orientation and the position of segment j change arbitrarily. Segment i and segment k rotate arbitrarily around their corresponding joint axis. \par
\begin{figure}[!ht]
\vspace{-2 mm}
\includegraphics[width=1\linewidth]{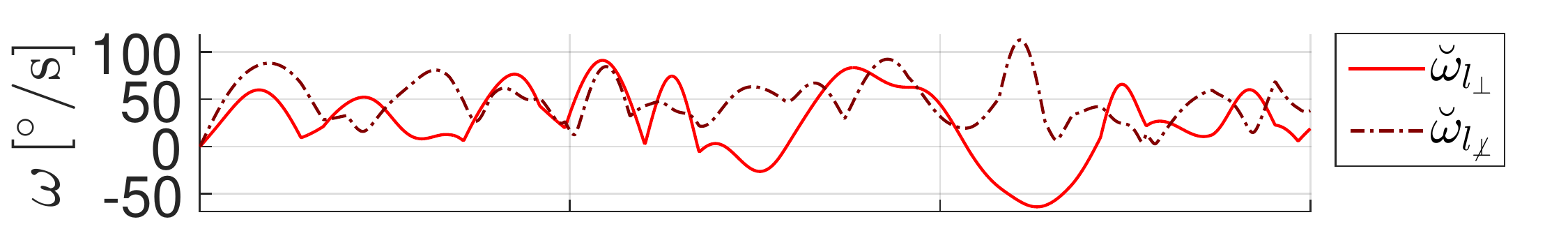}  
\includegraphics[width=1\linewidth]{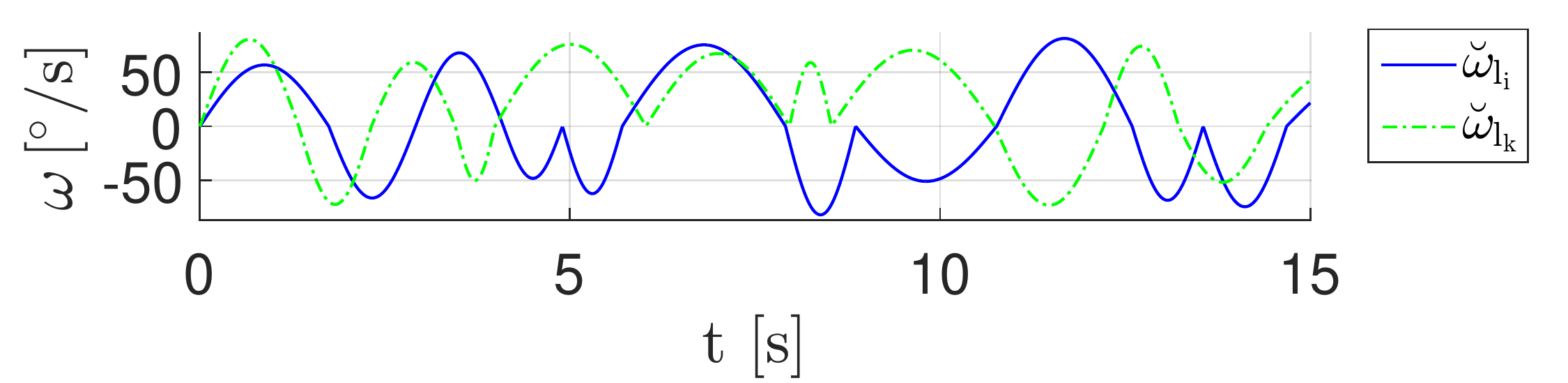}  
\caption{Angular velocities $\angvelParJ$, $\angvelResJ$, $\angvelJI$ and $\angvelJK$ projected onto their corresponding axis for the random movement.}
\label{fig:rdM}
\end{figure}
Notice that $\angvelParJP$ and $\angvelResJP$ are non-zero for almost all time instances.

\subsection{Results}
We compare the orientations estimated by the MHE to the true known orientations of the segments. To quantify the disagreement between a true and an estimated orientation, we use a metric called angular distance defined in~\cite{Hartley:2013}. It quantifies the smallest angle by which an orientation must be rotated to become identical to another orientation. We determine this orientation error for the relative orientation between segment i and j as well as for the relative orientation between j and k, and we denote these errors $\quatRelJErrAng$ and $\quatRelKErrAng$. If both errors are close to zero, then all orientations are well estimated and observability is confirmed in the sense of Theorem~\ref{th:obsth} and in the sense of Corollary~\ref{cor:obs}. \par
\begin{figure}[!ht]
\vspace{-2 mm}
\includegraphics[width=1\linewidth]{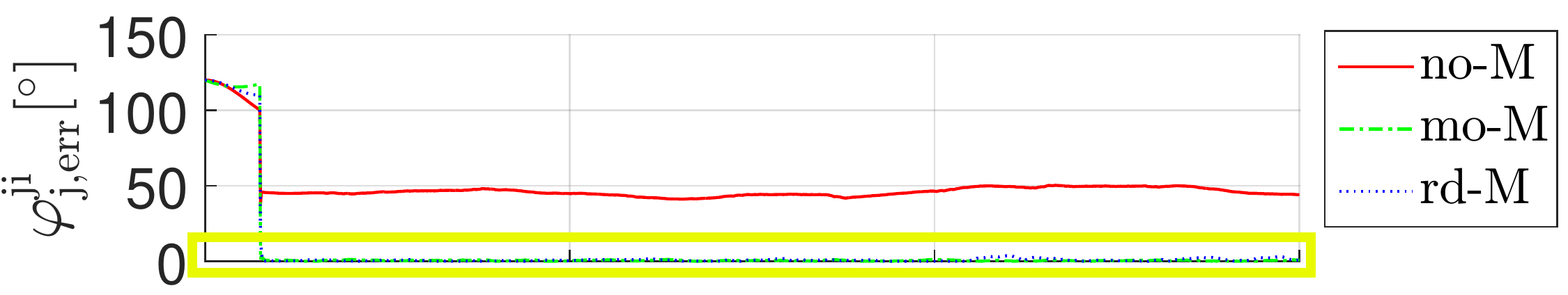}  
\includegraphics[width=1\linewidth]{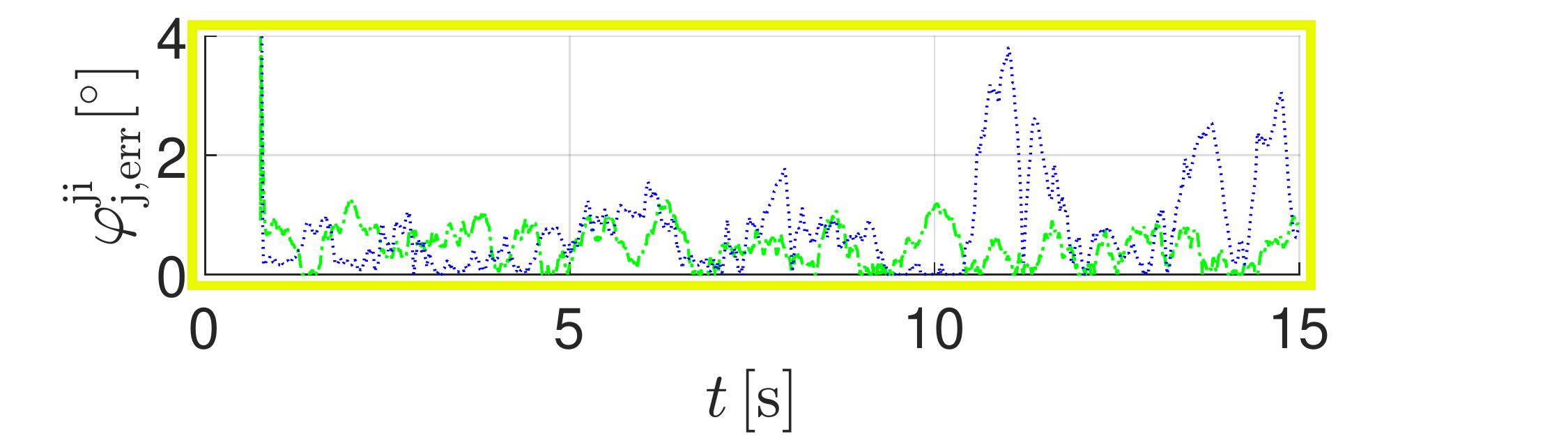}  
\caption{Relative orientation error $\quatRelJErrAng$ between the orientation $\quatI$ of segment i and the orientation $\quatJ$ of segment j.}
\label{fig:relOriJ}
\end{figure}
\begin{figure}[!ht]
\vspace{-2 mm}
\includegraphics[width=1\linewidth]{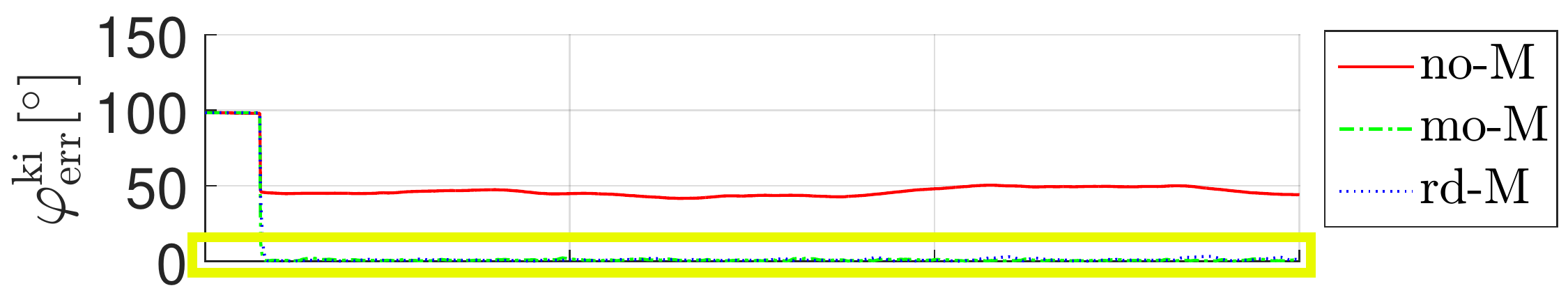} 
\includegraphics[width=1\linewidth]{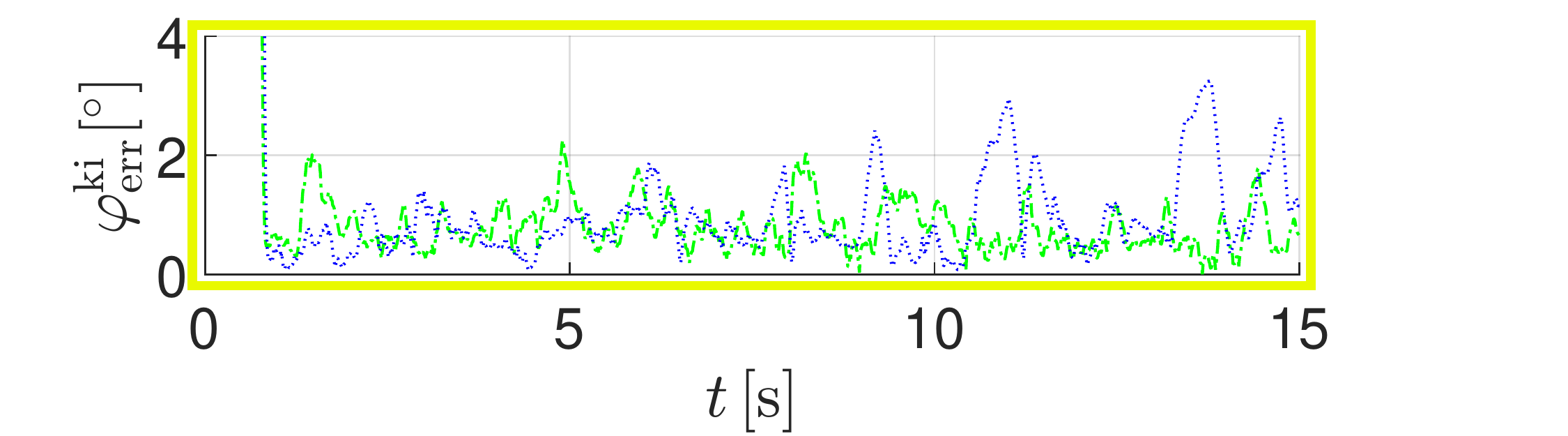}  
\caption{Relative orientation error $\quatRelKErrAng$ between the orientation $\quatI$ of segment i and the orientation $\quatK$ of segment k.}
\label{fig:relOriK}
\end{figure}

We determine both errors for all three motions (no-M, mo-M, rd-M) and both estimation modes (m1, m2). As expected, modes m1 and m2 yield equivalent results for all motions. Therefore, we only present results for the mode m2, in which none of the orientations is known a-priori. Fig.~\ref{fig:relOriJ} and Fig.~\ref{fig:relOriK} show the relative orientation errors $\quatRelJErrAng$ and $\quatRelKErrAng$ plotted over time. 
For the minimal-observable and the random movement, the relative orientation errors decrease rapidly to less than $4^\circ$. This agrees with Theorem~\ref{th:obsth}, because for both movements the angular velocities $\angvelParJP$ and $\angvelResJP$ are non-zero for almost all time instances, which implies that the relative orientations are instantaneously observable under partially unknown input for almost all time instances.
For the non-observable movement, the relative orientation errors remain as large as $45^\circ$ and drift over time, i.e. the relative orientations $\quatRel$ and $\quatRelK$ do not converge to the true relative orientations. This result is in agreement with Remark~\ref{rm:notobs}, because the angular velocity $\angvelParJP$ is constantly zero, which implies that the relative orientations are not instantaneously observable under partially unknown input.\par

\section{Conclusion}
We considered the task of inertial real-time motion tracking of kinematic joints and addressed two major challenges simultaneously: using a sparse sensor setup and avoiding magnetometer readings. Nonlinear kinematic joint constraints were exploited to overcome these challenges. While previous research on constraint-exploiting approaches commonly states that ``sufficient and persistent excitation of the degrees of freedom of the joints'' is required, we thoroughly analyzed what this means. For a double-hinge joint system, we proposed simple criteria which assure that the rotational motion states are observable from sparse and magnetometer-free inertial measurements.  \par

Furthermore, we proposed a moving-horizon estimation approach that was validated in a simulation study for three motions with different levels and qualities of excitation. All simulation results were in agreement with the theoretical conjectures. It was demonstrated that motions with strong excitation of many degrees of freedom of a kinematic chain might nevertheless lead to unobservable orientations, while some motions that excite only very few degrees of freedom might lead to instantaneous observability and to small estimation errors even in the presence of bias and noise errors. This emphasizes the value of precise observability conditions over  general demands for ``sufficient excitation''. \par 

All results apply likewise to joints with two rotational degrees of freedom, such as the human elbow or ankle joint, since they can be seen as double-hinge joints with short (or even zero-length) middle segments. Future research will aim at experimental validation and at transferring the new approach to additional constraints and other joint types and combinations.

\par
Future work might focus on the observability analysis of the case of parallel joint axes or different joint systems.

\bibliography{ifacconf}             

\begin{thebibliography}{21}
\providecommand{\natexlab}[1]{#1}
\providecommand{\url}[1]{\texttt{#1}}
\providecommand{\urlprefix}{URL }
\expandafter\ifx\csname urlstyle\endcsname\relax
  \providecommand{\doi}[1]{doi:\discretionary{}{}{}#1}\else
  \providecommand{\doi}{doi:\discretionary{}{}{}\begingroup
  \urlstyle{rm}\Url}\fi

\bibitem[{Andersson et~al.(2019)Andersson, Gillis, Horn, Rawlings, and
  Diehl}]{Anderson2018}
Andersson, J.A., Gillis, J., Horn, G., Rawlings, J.B., and Diehl, M. (2019).
\newblock Casadi: a software framework for nonlinear optimization and optimal
  control.
\newblock \emph{Mathematical Programming Computation}, 11(1), 1--36.

\bibitem[{Besan{\c{c}}on(2007)}]{besancon:2007}
Besan{\c{c}}on, G. (2007).
\newblock \emph{Nonlinear observers and applications}, volume 363.
\newblock Springer.

\bibitem[{Buke et~al.(2015)Buke, Gaoli, Yongcai, Lei, and Zhiqi}]{Buke:2015}
Buke, A., Gaoli, F., Yongcai, W., Lei, S., and Zhiqi, Y. (2015).
\newblock Healthcare algorithms by wearable inertial sensors: a survey.
\newblock \emph{China Communications}, 12(4), 1--12.

\bibitem[{Girrbach et~al.(2017)Girrbach, Hol, Bellusci, and
  Diehl}]{Girrbach:2017}
Girrbach, F., Hol, J., Bellusci, G., and Diehl, M. (2017).
\newblock Optimization-based sensor fusion of gnss and imu using a moving
  horizon approach.
\newblock \emph{Sensors}, 17(5), 1159.

\bibitem[{Graurock et~al.(2016)Graurock, Schauer, and
  Seel}]{Graurock2016_IFESS}
Graurock, D., Schauer, T., and Seel, T. (2016).
\newblock User-adaptive inertial sensor network for feedback-controlled gait
  support systems.
\newblock In \emph{Proc. of the 20th Annual International FES Society
  Conference}, 1–4. La Grande Motte, France.

\bibitem[{Hartley et~al.(2013)Hartley, Trumpf, Dai, and Li}]{Hartley:2013}
Hartley, R., Trumpf, J., Dai, Y., and Li, H. (2013).
\newblock Rotation averaging.
\newblock \emph{International journal of computer vision}, 103(3), 267--305.

\bibitem[{HSL(2007)}]{HSL}
HSL (2007).
\newblock A collection of fortran codes for large-scale scientific computation.
\newblock \emph{See http://www.hsl.rl.ac.uk}.

\bibitem[{Huang et~al.(2018)Huang, Kaufmann, Aksan, Black, Hilliges, and
  Pons-Moll}]{Huang:2018}
Huang, Y., Kaufmann, M., Aksan, E., Black, M., Hilliges, O., and Pons-Moll, G.
  (2018).
\newblock Deep inertial poser: Learning to reconstruct human pose from sparse
  inertial measurements in real time.
\newblock \emph{ACM Transactions on Graphics}, 37, 1--15.

\bibitem[{Kok et~al.(2014)Kok, Hol, and Sch{\"o}n}]{kokHS:2014}
Kok, M., Hol, J.D., and Sch{\"o}n, T.B. (2014).
\newblock An optimization-based approach to human body motion capture using
  inertial sensors.
\newblock In \emph{Proc. of the 19th World Congress of the International
  Federation of Automatic Control}, 79--85. Cape Town, South Africa.

\bibitem[{Kok et~al.(2017)Kok, Hol, and Sch{\"o}n}]{kokHS:2017}
Kok, M., Hol, J.D., and Sch{\"o}n, T.B. (2017).
\newblock Using inertial sensors for position and orientation estimation.
\newblock \emph{Foundations and Trends on Signal Processing}, 11(1-2), 1--153.

\bibitem[{Laidig et~al.(2019)Laidig, Lehmann, and Seel}]{laidigLS:2019}
Laidig, D., Lehmann, D., and Seel, T. (2019).
\newblock Magnetometer-free realtime inertial motion tracking by exploitation
  of kinematic constraints in 2-dof joints.
\newblock In \emph{41st IEEE International Engineering in Medicine and Biology
  Conference (EMBC)}. Berlin, Germany.

\bibitem[{Laidig et~al.(2017{\natexlab{a}})Laidig, M{\"u}ller, and
  Seel}]{laidigMS:2017}
Laidig, D., M{\"u}ller, P., and Seel, T. (2017{\natexlab{a}}).
\newblock Automatic anatomical calibration for imu-based elbow angle
  measurement in disturbed magnetic fields.
\newblock \emph{Current directions in biomedical engineering}, 3(2), 167--170.

\bibitem[{Laidig et~al.(2017{\natexlab{b}})Laidig, Schauer, and
  Seel}]{laidigSS:2017}
Laidig, D., Schauer, T., and Seel, T. (2017{\natexlab{b}}).
\newblock Exploiting kinematic constraints to compensate magnetic disturbances
  when calculating joint angles of approximate hinge joints from orientation
  estimates of inertial sensors.
\newblock In \emph{2017 International Conference on Rehabilitation Robotics
  (ICORR)}, 971--976. London, UK.

\bibitem[{Lerner(1978)}]{lerner:1978}
Lerner, G.M. (1978).
\newblock Three-axis attitude determination.
\newblock In J.R. Wertz (ed.), \emph{Spacecraft Attitude Determination and
  Control}, chapter 12.2, 421. Springer Netherlands.

\bibitem[{Lucia et~al.(2017)Lucia, T{\u{a}}tulea-Codrean, Schoppmeyer, and
  Engell}]{LuciaTSE:2017}
Lucia, S., T{\u{a}}tulea-Codrean, A., Schoppmeyer, C., and Engell, S. (2017).
\newblock Rapid development of modular and sustainable nonlinear model
  predictive control solutions.
\newblock \emph{Control Engineering Practice}, 60, 51--62.

\bibitem[{Miezal et~al.(2016)Miezal, Taetz, and Bleser}]{Miezal:2016}
Miezal, M., Taetz, B., and Bleser, G. (2016).
\newblock On inertial body tracking in the presence of model calibration
  errors.
\newblock \emph{Sensors}, 16, 1132.

\bibitem[{Olsson et~al.(2019)Olsson, Seel, Lehmann, and
  Halvorsen}]{Olsson2019_FUSION}
Olsson, F., Seel, T., Lehmann, D., and Halvorsen, K. (2019).
\newblock Joint axis estimation for fast and slow movements using weighted
  gyroscope and acceleration constraints.
\newblock In \emph{22nd International Conference on Information Fusion
  (FUSION)}. Ottawa, Canada.

\bibitem[{Taetz et~al.(2016)Taetz, Bleser, and Miezal}]{taetzBM:2016}
Taetz, B., Bleser, G., and Miezal, M. (2016).
\newblock Towards self-calibrating inertial body motion capture.
\newblock In \emph{Proceedings of the 19th International Conference on
  Information Fusion}, 1751--1759. Heidelberg, Germany.

\bibitem[{von Marcard et~al.(2017)von Marcard, Rosenhahn, Black, and
  Pons-Moll}]{Marcard:2017}
von Marcard, T., Rosenhahn, B., Black, M.J., and Pons-Moll, G. (2017).
\newblock Sparse inertial poser: Automatic 3d human pose estimation from sparse
  imus.
\newblock \emph{Comput. Graph. Forum}, 36, 349--360.

\bibitem[{W{\"a}chter and Biegler(2006)}]{WaechterB:2006}
W{\"a}chter, A. and Biegler, L. (2006).
\newblock On the implementation of an interior-point filter line-search
  algorithm for large-scale nonlinear programming.
\newblock \emph{Mathematical programming}, 106, 25--57.

\bibitem[{Wong et~al.(2015)Wong, Zhang, Lo, and Yang}]{Wong:2015}
Wong, C., Zhang, Z.Q., Lo, B., and Yang, G.Z. (2015).
\newblock Wearable sensing for solid biomechanics: A review.
\newblock \emph{IEEE Sensors Journal}, 15(5), 2747--2760.

\end{thebibliography}
                                                   







\appendix
\section{Proof of Theorem \ref{th:obsth}}\label{app:proof}
The following lemma, which is based on \cite{lerner:1978}, will be used in the proof.
\begin{lem}
If $\bm{v}$ and $\bm{w}$ are two non-parallel vectors in $\mathbb{R}^3$, then there exists only one unique orientation $\bm{R}^\mathrm{f}_\mathrm{g}$ that fulfills $\bm{v}^\mathrm{f}=\bm{R}^\mathrm{f}_\mathrm{g}\bm{v}^\mathrm{g}$ and $\bm{w}^\mathrm{f}=\bm{R}^\mathrm{f}_\mathrm{g}\bm{w}^\mathrm{g}$. If $\bm{v}$ and $\bm{w}$ are parallel, infinitely many such rotation matrices exist.
\label{lem:wahba}
\end{lem}
\begin{pf}
As a first step we show that there exist a mapping \eqref{eq:outmapp} that does not contain the unknown input $\cYgyrJ$. For $\bm{y}_1$ and $\bm{y}_2$ we choose an order of zero and for $y_3$ we also consider it's first derivative.
The derivative of $y_3$ yields
\begin{small}
\begin{subequations}
\begin{align}
\dot{y}_3 &= \frac{d}{dt}\left(\left( \cRIn \cYgyrI - \cRKn \cYgyrK \right)^\Transp \left( \cRIn \jointAxisI{i} \times \cRKn \jointAxisK{k} \right)\right)
\\
&= \left( \cRIn \cYgyrIdot - \cRKn \cYgyrKdot \right)^\Transp \left( \cRIn \jointAxisI{i} \times \cRKn \jointAxisK{k} \right)
\nonumber\\
& \quad + \left( \cRIn \cYgyrI - \cRKn \cYgyrK \right)^\Transp 
\nonumber\\
&\quad \left( \cRIn \crossP{\cYgyrI} \jointAxisI{i} \times \cRKn \jointAxisK{k} + \cRIn \jointAxisI{i} \times \cRKn \crossP{\cYgyrK} \jointAxisK{k} \right),
\end{align}
\end{subequations}
\end{small}
which shows that the chosen mapping does not contain the unknown input $\cYgyrJ$.\par
Now we proof that the orientations $\cRIn$ and $\cRKn$ are uniquely defined by the system of equations $y_3=0$ and $\dot{y}_3=0$ if the stated conditions in Theorem \ref{th:obsth} are fulfilled.\par
Partition the angular velocities of segment i and segment k into the angular velocity $\cYgyrJ$ of segment j and the change of the corresponding joint angle $\angvelJI$/$\angvelJK$:
\begin{subequations}
\begin{align}
\cYgyrI &= \cRIJ \cYgyrJ + \underbrace{\cYgyrI - \cRIJ \cYgyrJ}_{\angvelJI},
\\
\cYgyrK &= \cRKJ \cYgyrJ + \underbrace{\cYgyrK - \cRKJ \cYgyrJ}_{\angvelJK}.
\end{align}
\label{eq:partitioningIK}%
\end{subequations}
Further partition the angular velocity of segment j into the portion $\angvelParJ$ around the axis $\jointAxisPer$ and the residue $\angvelResJ$
\begin{align}
\cYgyrJ =& \overbrace{\left. \cYgyrJ \right.^\Transp \jointAxisPerJ \jointAxisPerJ}^{\angvelParJ} + \underbrace{\cYgyrJ - \left. \cYgyrJ \right.^\Transp \jointAxisPerJ \jointAxisPerJ}_{\angvelResJ},
\label{eq:partitioningJ}
\end{align}
where $\jointAxisPerJ = \frac{\jointAxisI{j} \times \jointAxisK{j}}{\norm{\jointAxisI{j} \times \jointAxisK{j}}}$.\par
If we now apply the two partitionings \eqref{eq:partitioningIK} and \eqref{eq:partitioningJ} to the constraint \eqref{eq:HJconst2d} and cancel every portion of the angular velocity which is perpendicular to the axis $\jointAxisPer$, we end up with equation
\begin{align}
y_3 = \Big( \underbrace{ \cRIn \cRIJ \angvelParJ - \cRKn \cRKJ \angvelParJ }_{\bm{\Omega}} \Big)^\Transp \jointAxisPerR = 0\label{eq:y3}
\end{align}
The scalar product of two parallel vectors can only be zero if one of the vectors is zero. The second argument of the scalar product is by definition unequal to zero, hence $\bm{\Omega}$ must be zero to guarantee that $y_3$ is zero.
\begin{align}
\dot{y}_3 &= \left[\frac{d}{dt} \bm{\Omega} \right]^\Transp \jointAxisPerR + \bm{\Omega}^\Transp \left[\frac{d}{dt} \jointAxisPerR \right] = 0\label{eq:y3dot}
\end{align}
We conclude that $y_3 = \dot{y}_3 = 0$ implies $\bm{\Omega} = \frac{d}{dt}\bm{\Omega} = \bm{0}$.\par
The derivative of $\bm{\Omega}$ yields:
\begingroup
\allowdisplaybreaks
\begin{small}
\begin{subequations}
\begin{align}
\frac{d}{dt} \bm{\Omega} &= \frac{d}{dt} \left( \cRIn \cRIJ \angvelParJ - \cRKn \cRKJ \angvelParJ \right)
\label{eq:constDer1}\\[5pt]
&=  \cRIndot \cRIJ \angvelParJ + \cRIn \cRIJdot \angvelParJ + \cRIn \cRIJ \angvelParJdot 
\nonumber\\
& \quad - \left( \cRKndot \cRKJ \angvelParJ + \cRKn \cRKJdot \angvelParJ + \cRKn \cRKJ \angvelParJdot \right)
\label{eq:constDer2}\\[5pt]
&= \bigg( \cRIn \crossP{\cYgyrI} \cRIJ \angvelParJ 
\nonumber\\
& \quad + \cRIn \left( -\crossP{\cYgyrI} \cRIJ + \cRIJ \crossP{\cYgyrJ} \right) \angvelParJ \bigg)
\nonumber\\
& \quad - \bigg( \cRKn \crossP{\cYgyrK} \cRKJ \angvelParJ 
\nonumber\\
& \quad + \cRKn \left( -\crossP{\cYgyrK} \cRKJ + \cRKJ \crossP{\cYgyrJ} \right) \angvelParJ \bigg)
\label{eq:constDer3}\\[5pt]
&= \bigg( \cRIn \crossP{ \cRIJ \cYgyrJ + \cYgyrI - \cRIJ \cYgyrJ } \cRIJ \angvelParJ
\nonumber\\
& \quad + \cRIn \left( -\crossP{\cYgyrI} \cRIJ + \cRIJ \crossP{\cYgyrJ} \right) \angvelParJ \bigg)
\nonumber\\
& \quad - \bigg( \cRKn  \crossP{ \cRKJ \cYgyrJ + \cYgyrK - \cRKJ \cYgyrJ } \cRKJ \angvelParJ
\nonumber\\
& \quad + \cRKn \left( -\crossP{\cYgyrK} \cRKJ + \cRKJ \crossP{\cYgyrJ} \right) \angvelParJ \bigg)
\label{eq:constDer4}\\[5pt]
&=  \cRIn \cRIJ \crossP{ \angvelResJ } \angvelParJ - \cRKn \cRKJ \crossP{ \angvelResJ } \angvelParJ. \label{eq:constDer5}
\end{align}
\end{subequations}
\end{small}
\endgroup
Here, we used the state dynamics \eqref{eq:HJsys_dyn} to advance from \eqref{eq:constDer2} to \eqref{eq:constDer3}, the partitioning \eqref{eq:partitioningIK} to advance from \eqref{eq:constDer3} to \eqref{eq:constDer4} and the partitioning \eqref{eq:partitioningJ} to turn \eqref{eq:constDer4} into \eqref{eq:constDer5}. \par
Therefore, $\bm{\Omega} = \bm{0}$ and $\frac{d}{dt}\bm{\Omega} = \bm{0}$ form the following system of equations:
\begin{subequations}
\begin{align}
\overbrace{\cRIJ \angvelParJ}^{\bm{v}^\mathrm{b_i}} =& \cRIK \overbrace{\cRKJ \angvelParJ}^{\bm{v}^\mathrm{b_k}}
\\
\underbrace{\cRIJ \left( \angvelResJ \times \angvelParJ \right)}_{\bm{w}^\mathrm{b_i}} =& \cRIK \underbrace{\cRKJ \left( \angvelResJ \times \angvelParJ \right)}_{\bm{w}^\mathrm{b_k}}
\end{align}
\label{eq:wahbas}
\end{subequations}
Applying Lemma~\ref{lem:wahba} to \eqref{eq:wahbas} allows us to state the following. There exists only one unique orientation $\cRIK$ that fulfills \eqref{eq:wahbas} if the vectors $\bm{v}$ and $\bm{w}$ are linearly independent. Hence the following two conditions need to be fulfilled:
\begin{itemize}
\item The angular velocity $\angvelParJ$ of segment j around the axis $\jointAxisPer$ is unequal to zero.
\item The residue $\angvelResJ$ of the angular velocity of segment j is unequal to zero.
\end{itemize}
Since every solution of $y_3=\dot{y}_3=0$ must be a solution of \eqref{eq:wahbas}, there exists only one unique relative orientation $\cRIK$ that fulfills $y_3=\dot{y}_3=0$ if the stated conditions are fulfilled.

If additionally one of the orientations $\cRIn$ and $\cRKn$ is known, the 
other can be determined using the relative orientation $\cRIK$. \par

If the orientations of segment i and segment k are known Lemma~\ref{lem:wahba} can be applied to $\bm{y}_1=\bm{y}_2=0$ to proof that the orientation $\cRJn$ is uniquely defined by the system of equations $\bm{y}_1=0$ and $\bm{y}_2=0$.\par
Hence the mapping is injective.
\qed 
\end{pf}
\end{document}